\documentclass[12pt]{article}

\title{{\Large Two-qubit catalysis in a four-state pure bipartite system}}
\author{Peter H. Anspach \\ {\sl National Security Agency, Suite 6409 } \\ {\sl 9800 Savage Road} \\ {\sl Fort George G. Meade, MD 20755}}
\begin{document}
\maketitle
\begin{abstract}
We consider a four-state pure bipartite system consisting of four qubits
shared among two parties using the same Schmidt basis, two qubits per party.
In some cases, transformation between two known pure states may not be
possible using LOCC transformations but may be possible with the addition
of a two-qubit catalyst.  We provide a necessary and sufficient condition
for this to occur.
\end{abstract}

\section{Introduction}

Consider the case of two people, Alice and Bob, who live far apart. Some time
ago, Alice boarded a plane with some of her qubits, visited Bob in
his laboratory, entangled her qubits with his, and returned home with her
qubits.  Together, their qubits form the known pure state $|\psi\rangle$.  Now
they wish to perform an experiment, but one which requires the joint
system to be in the known pure state $|\phi\rangle$.  It would be very 
inconvenient for Alice to fly back to Bob's lab, but it is easy for her
to phone him.  It would be nice if they could change the state 
$|\psi\rangle$ to the state $|\phi\rangle$ by each of them operating
locally on their portion of the system and exchanging classical information
as necessary.  Such a transformation is called an LOCC
transformation (Local Operations and Classical Communication).  Note that
the local operations are not necessarily unitary nor are they confined to 
the qubits which
form the entangled state.  Both Alice and Bob may bring in ancilla
qubits in various states of entanglement and entangle them with their
local systems.  However they do not have any ancillary quantum
communication channels.  In summary,
Alice and Bob may perform arbitrary local
operations on their local systems, but may only communicate using 
classical communication channels.

Majorization is a useful concept which relates to convexity.  Given a
vector $\vec\alpha$, let $\alpha_{[1]}$ be the largest component of the
vector, $\alpha_{[2]}$ the second largest component, and so on.  The
n-long vector $\vec\alpha$ is majorized by the n-long vector $\vec\alpha'$ 
iff
$$\sum_{i=1}^k\alpha_{[i]}\leq \sum_{i=1}^k\alpha_{[i]}'\quad k=1,\ldots,n-1$$
and
$$\sum_{i=1}^n\alpha_{[i]}= \sum_{i=1}^n\alpha_{[i]}'.$$

Nielsen's Theorem {\bf [1]} gives a necessary and sufficient condition for an
LOCC to be possible.  We will 
assume that all states have the same basis for their Schmidt decompositions.
This assumption is not necessary for obtaining any of the results, but it
simplies the notation and computation enormously. 
With this assumption, Nielsen's Theorem reduces to the following:

\vspace{.125in}
{\it
\noindent Let
$|\psi\rangle = \sum_{i=1}^n \sqrt{\alpha_i}\,\, |i_A\rangle |i_B\rangle$
and
$|\phi\rangle = \sum_{i=1}^n \sqrt{\alpha_i'}\,\, |i_A'\rangle |i_B'\rangle$
be pure bipartite states with respective Schmidt coefficients 
$\alpha_1\geq\alpha_2\geq\ldots\alpha_n\geq0$ 
and
$\alpha_1'\geq\alpha_2'\geq\ldots\alpha_n'\geq0$.
Then the transformation $|\psi\rangle\to|\phi\rangle$ can be performed
under LOCC iff the vector $\vec\alpha$ is majorized by the vector
$\vec\alpha'$.
}
\vspace{.125in}

\section{Catalysis}

There are clearly cases where LOCC transformations cannot occur, {\it i.e.}
when $\vec\alpha$ is not majorized by $\vec\alpha'$.  However all is not
necessarily lost for Alice and Bob; there is still the possibility of
catalysis.  Suppose that the Quantum Entanglement Savings Bank has branches
in both Alice's and Bob's home town.  The branch in Alice's town has a qubit 
which is entangled with a qubit held by the branch in Bob's town; we will
assume that this is also a pure bipartite state of the form
$|\kappa\rangle = \sqrt{p}\,\, |00\rangle + \sqrt{1-p}\,\, |11\rangle$.  Alice
and Bob may borrow these qubits from the bank; however, the bank demands
that the qubits be returned to them in exactly the same state $|\kappa\rangle$.
(How exactly the bankers manage to verify this is not our problem.)

At first, this seems to be of no help.  Certainly if Alice and Bob were allowed
to tamper with the entanglement of $|\kappa\rangle$, they could use this
to transmit quantum information and thus perform a larger class of operations.
But such a process would reduce the Von Neumann entropy of the pair.
Requiring the state $|\kappa\rangle$ at the end of the process effectively means
that there can be no net transfer of entanglement from these ancillary 
particles to the
original system (however one chooses to measure entanglement).  Nevertheless,
the resource of entanglement can be borrowed provided it is returned at the
end.

In {\bf [2]}, Johnathan and Plenio examine the case when $n=4$.  They show
that the only case in which a transformation cannot be performed under
LOCC but can be performed under LOCC with a catalyst is when
$$\alpha_1\leq\alpha_1',\quad\alpha_1+\alpha_2>\alpha_1'+\alpha_2',\quad
\alpha_4\geq\alpha_4'\leqno{(*)}$$
and they provide an example:
$$\begin{array}{rlllr}
|\psi\rangle &= \sqrt{0.4}\,\,|00\rangle &+\sqrt{0.4}\,\,|11\rangle &+\sqrt{0.1}
|22\rangle &+ \sqrt{0.1}\,\,|33\rangle \\
|\phi\rangle &= \sqrt{0.5}\,\,|00\rangle &+\sqrt{0.25}\,\,|11\rangle &+\sqrt{0.25}\,\,|22\rangle &+ 0\,\, |33\rangle\\
\end{array}$$
They show the transformation cannot be performed under LOCC, but can be
done with the addition of the catalyst
$$|\kappa\rangle = \sqrt{0.6}\,\,|00\rangle + \sqrt{0.4}\,\,|11\rangle.$$

To verify this example, we will note the following fact:
If we let $\beta_i$
and $\beta_i'$ be the pure bipartite state coefficients of the augmented 
systems $|\psi\rangle|\kappa\rangle$ and $|\phi\rangle|\kappa\rangle$, then
$\{\beta_i\}=\{\alpha_i p, \alpha_i (1-p)\}$ and 
$\{\beta_i'\}=\{\alpha_i' p, \alpha_i' (1-p)\}$.  By Nielsen's Theorem, 
a catalytic conversion is possible iff $\vec\beta$ is majorized by
$\vec\beta'$. It is thus easy to calculate the components of the 
augmented systems and verify that majorization occurs.

The proof of $(*)$ is fairly easy and straightforward.  We will paraphrase
Jonathan and Plenio's proof here.  
Suppose transformation under catalysis is
possible.  Let $K$ be the largest component of the catalyst, and $k$ the
smallest component.  Then
the first partial sums of the new source and target states will be
$K\alpha_1$ and $K\alpha_1'$.  Because LOCC is now possible, we have
majorization and $K\alpha_1\leq K\alpha_1'$ and hence $\alpha_1\leq\alpha_1'$.
Similarly, the penultimate partial sums will be $1-k\alpha_4$ and 
$1-k\alpha_4'$.  Since majorization occurs, we have 
$1-k\alpha_4\leq 1-k\alpha_4'$ and hence $\alpha_4\geq\alpha_4'$.
Now, we are also supposing that transformation is not possible
under LOCC without the presence of a catalyst.  Hence we must have 
$\alpha\not\prec\alpha'$.  We have just shown $\alpha_1\leq\alpha_1'$.
Since $\alpha_4\geq\alpha_4'$, we have 
$$\alpha_1+\alpha_2+\alpha_3 = 1-\alpha_4 \leq 1-\alpha_4' = \alpha_1' 
+ \alpha_2' + \alpha_3'.$$  Finally, the $\alpha_i$'s and $\alpha_i'$'s 
must sum to 1, so the fourth partial sums are equal.  The only way
majorization can fail to occur is in the second partial sum.  Hence, we
must have $\alpha_1+\alpha_2>\alpha_1'+\alpha_2'$.  This completes the
proof.

Thus, when in the $n=4$ case, $(*)$ is a necessary 
condition
for catalysis to be effective with a pure bipartite two-qubits catalyst.
However, if we restrict ourselves to catalysts of this form, the condition
is not both necessary and sufficient.  Zhou and Guo {\bf [3]} give an
example of a five-state system in which no two-qubit catalyst can effect
an LOCC transformation.
Here we will give an example with a four-state system.
Consider the states
$$\begin{array}{rlllr}
|\psi\rangle &= \sqrt{0.45}\,\,|00\rangle &+\sqrt{0.45}\,\,|11\rangle &+
\sqrt{0.05}\,\,|22\rangle &+ \sqrt{0.05}\,\,|33\rangle \\
|\phi\rangle &= \sqrt{0.5}\,\,|00\rangle &+\sqrt{0.35}\,\,|11\rangle &+
\sqrt{0.15}\,\,|22\rangle &+ 0\,\, |33\rangle \\
\end{array}$$
They satisfies $(*)$; however if we apply a catalyst of the form
$$|\kappa\rangle = \sqrt{p}\,\, |00\rangle + \sqrt{1-p}\,\, |11\rangle,$$
the pure state bipartite coefficients will be
$$\begin{array}{rl}
\vec\beta &= (.45p,.45p,.05p,.05p,.45(1-p),.45(1-p),.05(1-p),.05(1-p)) \\
\vec\beta' &= (.50p,.35p,.15p,0,.50(1-p),.35(1-p),.15(1-p),0) \\
\end{array}$$
where $\beta$ and $\beta'$ are the vectors of eigenvalues of the joint
system obtained by pairing $|\psi\rangle$ and $|\phi\rangle$ respectively 
with the catalyst.
Let $\lambda_i$ be the sum of the 
$i$ largest eigenvalues in the vector $\vec\beta$ and $\lambda'_i$ be
the sum of the $i$ largest eigenvalues in the vector $\vec\beta'$.
The condition of $\vec\beta$ being majorized by $\vec\beta'$ then
becomes simply $\lambda_i\leq\lambda_i'$ for all $i$.

The two largest eigenvalues in $\vec\beta$ are both $.45p$ and hence
$\lambda_2 = .9p$.  Let us suppose $p>10/17$.  
Then $.35p>.50(1-p)$ and $\lambda_2'=.85p$.
Thus, $\lambda_2>\lambda_2'$ and we cannot perform an LOCC.  On the other
hand, suppose $1/2\leq p\leq10/17$.  In this case, the four largest 
eigenvalues of $\vec\beta$ are $.45p$, $.45p$, $.45(1-p)$, and $.45(1-p)$ 
yielding $\lambda_4 = .9$.  The four largest eigenvalues of $\vec\beta'$ are
$.50p$, $.50(1-p)$, $.35p$, and $.35(1-p)$ yielding $\lambda_4' = .85$.
Thus, $\lambda_4>\lambda_4'$ and again we cannot perform an LOCC.

This naturally raises the question ``What is a necessary and sufficient
condition to perform catalytic conversion under this set-up?''  We
have already seen that $(*)$ is necessary.  It is easy to see that 
$(*)$ is equivalent to the following:
$$\exists\, \epsilon_1\geq0,\epsilon_2>0,\epsilon_3\geq0\hbox{ such that}\leqno(**)$$
\vspace{-.25in}
$$\begin{array}{rl}
\alpha_1'&= \alpha_1 + \epsilon_1 \\ 
\alpha_2'&= \alpha_2 - \epsilon_1 - \epsilon_2 \\
\alpha_3'&= \alpha_3 + \epsilon_2 + \epsilon_3 \\
\alpha_4'&= \alpha_4 - \epsilon_3. \\
\end{array}$$

\noindent In illustration, let us look back at Jonathan and Plenio's example:
$$\begin{array}{rlllr}
|\psi\rangle &= \sqrt{0.4}\,\,|00\rangle &+\sqrt{0.4}\,\,|11\rangle &+
\sqrt{0.1}\,\,|22\rangle &+ \sqrt{0.1}\,\,|33\rangle \\
|\phi\rangle &= \sqrt{0.5}\,\,|00\rangle &+\sqrt{0.25}\,\,|11\rangle &+
\sqrt{0.25}\,\,|22\rangle &+ 0\,\, |33\rangle\\
&&&\\
|\kappa\rangle &= \sqrt{0.6}\,\,|00\rangle &+ \sqrt{0.4}\,\,|11\rangle &&$$
\end{array}$$
Here we have $\alpha_1 = .4$, $\alpha_2 = .4$, $\alpha_3 = .1$,
$\alpha_4 = .1$, $\epsilon_1 = .1$, $\epsilon_2 = .05$, and
$\epsilon_3 = .1$.  Also, $p=.6$.
\vspace{.125in}

{\it

\noindent {\bf Theorem:} Let
$$m={\rm max}\left({\alpha_2-\epsilon_1\over\alpha_1+\epsilon_1},
{\alpha_4 - \epsilon_3\over\alpha_3+\epsilon_3},
{\epsilon_2\over\epsilon_1}\right)$$
and
$$M={\rm min}\left({\alpha_3+\epsilon_3\over\alpha_2-\epsilon_1},
{\epsilon_3\over\epsilon_2}\right)$$
(We take $m=+\infty$ if $\epsilon_1=0$.)
The transformation $|\psi\rangle\to|\phi\rangle$ cannot
be performed under LOCC by itself but can be performed under LOCC
with a two-qubit pure bipartite state catalyst $|\kappa\rangle$ 
if and only if $(**)$ holds as above and $m\leq M$.  Moreover if this is
the case, then $|\kappa\rangle$ will be a valid catalyst for 
$|\psi\rangle\to|\phi\rangle$ iff
$$m\leq {1-p\over p} \leq M$$}
\vspace{.125in}

\section{The Proof}

The proof itself is rather cumbersome.  Because of this, we will give
a preliminary overview to help the reader follow the
logic.  We will start by using a few simple observations to reduce
the requirements of the proof.  We will then start out assuming 
that we are in the situation where catalysis occurs, and use this
fact to gain information about $\lambda$ and $\lambda'$.  It will
turn out that the condition of majorization forces only one possible
choice for the $\lambda_i'$'s, and we will then compute them.  There
will, unfortunately, be many choices for $\lambda$; however, each
$\lambda_i$ will be limited.  By breaking each one up into a few
cases, we can determine conditions that work for each case.  
(This is the bulk of the proof, Sections {\bf 3.5} through {\bf 3.12}.
They are, frankly, tedious to check and the reader may wish 
to skip them.)  It is useful to note that there is a certain symmetry
among the cases; in general, the $\lambda_i$ case is the mirror of
the $\lambda_{8-i}$ case.  (The $\lambda_8$ case itself is merely
$1=1$.)
Finally, we will show that the argument used for the
forward direction of the theorem is completely reversible and provides
a proof for the backward direction as well.

We will now begin the proof with a few preliminaries.  First, 
the quantity $(1-p)/p$
occurs frequently in our calculations; it will be convenient to refer
to it as $r$.  Note that $p=1/(1+r)$.  Since all the arguments of
$m$ are positive, $m\geq0$.  Since $\alpha_2'\geq\alpha_3'$, 
$\alpha_2-\epsilon_1-\epsilon_2\geq\alpha_3+\epsilon_2+\epsilon_3$, 
we have $\alpha_2-\epsilon_1\geq\alpha_3+\epsilon_3$.  Thus
$(\alpha_3+\epsilon_3)/(\alpha_2-\epsilon_1)\leq 1$.  This implies
$M\leq1$.  Therefore, if $m\leq r \leq M$, $0\leq r\leq 1$ and so
$1/2 \leq p \leq 1$.  Therefore, every
$r$ between $m$ and $M$ will produce a $p$ in the valid
range between $1/2$ and $1$.

Also, let us look at the vectors $\vec\beta$ and $\vec\beta'$.  
We have noted that the components of $\vec\beta$ are $\alpha_1 p$, 
$\alpha_2 p$, $\alpha_3 p$,
$\alpha_4 p$, $\alpha_1 (1-p)$, $\alpha_2 (1-p)$, $\alpha_3 (1-p)$ and
$\alpha_4 (1-p)$ and similarly for $\vec\beta'$.  We know
$\alpha_1\geq\alpha_2\geq\alpha_3\geq\alpha_4$ and $p\geq(1-p)$.  Therefore,
we have
$$\alpha_1 p\geq\alpha_2 p\geq\alpha_3 p\geq\alpha_4 p$$
$$\alpha_1 (1-p)\geq\alpha_2 (1-p)\geq\alpha_3 (1-p)\geq\alpha_4 (1-p).$$
We also know that $\alpha_i p \geq \alpha_i (1-p)$ for $i=1,\ldots,4$.
However, we do not {\it a~priori} have any additional knowledge about
the ordering of the components.

We will now prove a simple lemma:

\newpage
{\it

\noindent {\bf Lemma:} Suppose
$${a\over b}\geq{c\over d}.$$
Then
$${a\over b}\geq{a+c\over b+d}\geq{c\over d}.$$}
\vspace{.125in}

\noindent The proof is completely trivial.  Multiplying out the inequality
in the hypothesis yields $ad\geq bc$.  Adding $ab$ to both sides yields the
first inequality and adding $cd$ to both sides yields the second.  Despite
the simplicity of this lemma, it will frequently prove useful.

\vspace{.125in}

Our next observation is that the first part of the theorem follows
immediately from the second part.  If $m>M$, then it is impossible
to pick $r$ such that $m\leq r\leq M$.  Therefore there will be no
valid catalysts of the appropriate form and thus catalysis cannot
occur.  On the other hand, if $m\leq M$, we simply choose $r$
such that $m\leq r\leq M$.  In that case $|\kappa\rangle$ will allow catalytic
conversion to take place, and hence catalysis is possible.

To prove the second part of the theorem, fix two pure bipartite
states $|\psi\rangle$ and $|\phi\rangle$ which satisfy $(**)$ and fix
a catalyst $|\kappa\rangle$.  Let $m$, $M$, and $r$ be as specified above.
By Nielsen's Theorem, $|\kappa\rangle$ is a valid catalyst for an
LOCC transformation iff $\vec\beta$ is majorized by $\vec\beta'$.
This reduces the argument to proving the following statement: 
$\vec\beta$ is majorized by $\vec\beta'$ if and only if $m\leq r\leq M$. 

We begin with the forward direction:  Assume that $\vec\beta$ is majorized by 
$\vec\beta'$.  We will use this assumption to find a set of restrictions on
$r$.

\subsection{The first two components: $r\geq {\alpha_2'\over\alpha_1'}$}

Suppose $r< {\alpha_2'/\alpha_1'}$.  Then 
$\alpha_2' p > \alpha_1'(1-p)$, {\it i.e.}
$(\alpha_2-\epsilon_1-\epsilon_2) p> (\alpha_1+\epsilon_1) (1-p)$.  This implies
$\alpha_2 p > \alpha_1 (1-p)$.  We know that $\alpha_1 p$ is the largest
component of $\vec\beta$.  Since $\alpha_2 p > \alpha_1 (1-p)$, we know
that $\alpha_2 p$ is the second largest component.  Therefore, the
sum of the two largest components of $\vec\beta$ is 
$$\lambda_2 = \alpha_1 p + \alpha_2 p.$$
But we also know that $\alpha_2' p > \alpha_1'(1-p)$ and so the two largest
components $\vec\beta'$ are $\alpha_1' p$ and $\alpha_2' p$.
Thus
$$\begin{array}{rl}
\lambda_2' &= \alpha_1' p + \alpha_2' p \\
     &= \alpha_1 p + \epsilon_1 p + \alpha_2 p - \epsilon_1 p - \epsilon_2 p \\
           &= \alpha_1 p + \alpha_2 p - \epsilon_2 p \\
           &< \alpha_1 p + \alpha_2 p \\
           &= \lambda_2 \\
\end{array}$$
But $\vec\beta'$ majorizes $\vec\beta$ and so $\lambda_2'\geq\lambda_2$.  
This is a contradiction, so we must have 
$r\geq {\alpha_2'/\alpha_1'}$.

\subsection{The second and third components: $r\leq {\alpha_3'\over\alpha_2'}$}

Proof: Suppose $r > {\alpha_3'/\alpha_2'}$.  Then 
$\alpha_2'(1-p) > \alpha_3' p$.  From the proof of Step 1, we know the two
largest components of $\vec\beta'$ are $\alpha_1' p$, $\alpha_1' (1-p)$.
The component $\alpha_2' p$ is larger than any of the remaining components,
so it is the third largest.  The preceeding inequality shows that 
$\alpha_2' (1-p)$ is the fourth largest component.  Hence,
$$\lambda_4' = \alpha_1' + \alpha_2' = \alpha_1 + \alpha_2 - \epsilon_2.$$
Consider $\vec\beta$.  We know that $\alpha_1 p$, $\alpha_1 (1-p)$, 
$\alpha_2 p$, and $\alpha_2 (1-p)$ are four components of this vector.
If they are the four largest, then $\lambda_4 = \alpha_1 + \alpha_2$.  If
they are not the four largest, then $\lambda_4 > \alpha_1 + \alpha_2$.  In
either case, we have 
$$\begin{array}{rl}
\lambda_4 &\geq \alpha_1 + \alpha_2 \\
          &> \alpha_1 + \alpha_2 - \epsilon_2\\
          &= \lambda_4' \\
\end{array}$$
But again, $\vec\beta'$ majorizes $\vec\beta$ and so $\lambda_4'\geq\lambda_4$.
This is another contradiction, so we must have 
$r\leq {\alpha_3'/\alpha_2'}$.

\subsection{The last two components: $r\geq {\alpha_4'\over\alpha_3'}$}

Proof: This argument exactly mirrors the argument of {\bf 3.1}.  If the condition 
on $r$ were false, then $\alpha_3 (1-p)$ and 
$\alpha_4 (1-p)$ would be the smallest components of $\vec\beta$, and 
$\alpha_3' (1-p)$ and $\alpha_4' (1-p)$ the smallest components
of $\vec\beta'$.  Hence
$\lambda_6' = \alpha_1 + \alpha_2 + \alpha_3 p + \alpha_4 p -\epsilon_2 (1-p)$ and
$\lambda_6  = \alpha_1 + \alpha_2 +\alpha_3 p +\alpha_4 p $.  Once
again we would have $\lambda_6'<\lambda_6$, which cannot occur because
$\vec\beta'$ majorizes $\vec\beta$.  Therefore, we must have $r\geq \alpha_4'/\alpha_3'$

\subsection{Computing the $\lambda_i'$'s.}

We now know the ordering of the components of $\vec\beta'$ for valid
catalysts:  We know 
$\alpha_1' p\geq \alpha_1' (1-p)$ since $p\geq(1-p)$.  We have 
$\alpha_1' (1-p) \geq \alpha_2' p$ by {\bf 3.1}.  We know 
$\alpha_2' p \geq \alpha_3' p$ since $\alpha_2'\geq\alpha_3'$.  We have
$\alpha_3' p\geq \alpha_2' (1-p)$ by {\bf 3.2}.  We know 
$\alpha_2' (1-p) \geq \alpha_3' (1-p)$ since $\alpha_2'\geq\alpha_3'$.
We have $\alpha_3' (1-p) \geq \alpha_4' p$ by {\bf 3.3}.  And we have
$\alpha_4' p\geq\alpha_4' (1-p)$ since $p\geq(1-p)$.  This means we can
calculate the $\lambda_i'$'s:
$$\begin{array}{ll}
\lambda_1' &= \alpha_1 p + \epsilon_1 p \\
\lambda_2' &= \alpha_1 + \epsilon_1 \\
\lambda_3' &= \alpha_1 + \alpha_2 p + \epsilon_1 (1-p) - \epsilon_2 p \\
\lambda_4' &= \alpha_1 + \alpha_2 p + \alpha_3 p + \epsilon_1 (1-p) 
       + \epsilon_3 p \\
\lambda_5' &= \alpha_1 + \alpha_2 + \alpha_3 p - \epsilon_2 (1-p) 
       + \epsilon_3 p \\
\lambda_6' &= \alpha_1 + \alpha_2 + \alpha_3 + \epsilon_3 \\
\lambda_7' &= \alpha_1 + \alpha_2 + \alpha_3 + \alpha_4 p + \epsilon_3 (1-p) \\
\lambda_8' &= \alpha_1 + \alpha_2 + \alpha_3 + \alpha_4 = 1 \\
\end{array}$$

Since we are assuming that $\vec\beta$ is majorized by $\vec\beta'$, we know
that $\lambda_i'\geq\lambda_i$ for $i=1,\ldots,8$ with equality holding for
$i=8$.  We will analyze each of these inequalities and see what restrictions
they place on $r$.

\subsection{The first sum: $\lambda_1'\geq\lambda_1$ always holds.}

We know that $\alpha_1 p$ is the largest component in $\vec\beta$ so
$\lambda_1 = \alpha_1 p$.  Thus $\lambda_1' = \lambda_1 + \epsilon_1 p$ and
we have $\lambda_1'\geq\lambda_1$.  This inequality always holds and
there is no further restriction on $r$.

\subsection{The second sum: $\lambda_2'\geq\lambda_2$ iff 
$r\geq (\alpha_2-\epsilon_1)/(\alpha_1 + \epsilon_1)$.}

For this step, we will divide the proof into two cases.  First, suppose
that $r\leq\alpha_2/\alpha_1$.  This implies that 
$\alpha_1 (1-p)\leq\alpha_2 p$.  Thus $\lambda_2 = \alpha_1 p + \alpha_2 p$.
Therefore, the following inequalities are equivalent:
$$\begin{array}{rl}
\lambda_2'&\geq\lambda_2 \\
\alpha_1 + \epsilon_1&\geq \alpha_1 p + \alpha_2 p \\
(\alpha_1 + \epsilon_1)/(\alpha_1 + \alpha_2)&\geq p = {1/(1+r)} \\
1+r &\geq (\alpha_1 + \alpha_2)/(\alpha_1 + \epsilon_1)\\
r&\geq {(\alpha_2-\epsilon_1)/(\alpha_1 + \epsilon_1)} \\
\end{array}$$
which is the desired inequality.

For the second case,  suppose that $r>\alpha_2/\alpha_1$.  This implies that 
$\alpha_1 (1-p) >\alpha_2 p$.  Thus $\lambda_2 = \alpha_1$.  Since
$\lambda_2'=\alpha_1 + \epsilon_1$, we have $\lambda_2'\geq\lambda_2$.
But 
$$\alpha_2/\alpha_1>(\alpha_2-\epsilon_1)/(\alpha_1+\epsilon_1),$$
so $r>\alpha_2/\alpha_1$
implies $r\geq (\alpha_2-\epsilon_1)/(\alpha_1 + \epsilon_1)$.  Therefore
both inequalities are always true in this case.

We have therefore shown the desired equivalence.

\subsection{The third sum: $\lambda_3'\geq\lambda_3$ iff
$r\geq (\alpha_3+\epsilon_2)/(\alpha_1 + \epsilon_1)$ and
$r\geq \epsilon_2/\epsilon_1$.}

Again, we will split this into two cases.  First, suppose 
$r\leq\alpha_3/\alpha_1$.
This implies that $\alpha_1 (1-p) \leq \alpha_3 p$.  Then 
by the same reasoning as above, $\alpha_1 p$, $\alpha_2 p$, and
$\alpha_3 p$ are the three largest components of $\vec\beta$.  
Hence $\lambda_3 = \alpha_1 p + \alpha_2 p + \alpha_3 p$.
Therefore the following inequalities are equivalent:
$$\begin{array}{rl}
\lambda_3'&\geq\lambda_3 \\
\alpha_1 + \alpha_2 p + \epsilon_1 (1-p) - \epsilon_2 p 
&\geq \alpha_1 p + \alpha_2 p + \alpha_3 p\\
r&\geq {(\alpha_3+\epsilon_2)/(\alpha_1 + \epsilon_1)} \\
\end{array}$$
Notice also, that we have assumed $r\leq\alpha_3/\alpha_1$.  Since
$r\geq (\alpha_3+\epsilon_2)/(\alpha_1 + \epsilon_1)$, 
our preliminary lemma implies
$r\geq \epsilon_2/\epsilon_1$.  
Conversely, if
$r\geq (\alpha_3+\epsilon_2)/(\alpha_1 + \epsilon_1)$, then 
$\lambda_3'\geq\lambda_3$.

The second case, $r>\alpha_3/\alpha_1$, is the opposite of the
first.  Here $\lambda_3 = \alpha_1 + \alpha_2 p$. 
This leads to 
$\lambda_3'\geq\lambda_3 \iff
r\geq \epsilon_2/\epsilon_1$.
Also, we have assumed $r>\alpha_3/\alpha_1$ and shown
$r\geq \epsilon_2/\epsilon_1$, 
so by the preliminary lemma
$r\geq (\alpha_3+\epsilon_2)/(\alpha_1+\epsilon_1)$.  
Conversely, if $r\geq \epsilon_2/\epsilon_1$ and 
$r\geq (\alpha_3+\epsilon_2)/(\alpha_1 + \epsilon_1)$, then 
$\lambda_3'\geq\lambda_3$.

We have therefore shown the desired equivalence.  
\vspace{.125in}
\newline
\noindent {\bf Remark: }{\it Notice that this step
shows that $\epsilon_1>0$ is necessary for catalysis to occur.}

\subsection{The fourth sum: $\lambda_4'\geq\lambda_4$ iff
$r\geq (\alpha_4-\epsilon_3)/(\alpha_1 + \epsilon_1)$ and
$r\leq (\alpha_3+\epsilon_3)/(\alpha_2 - \epsilon_1)$.}

The argument for this is practically the same as Step 7, only we have three
cases. 
For the first case, suppose $r\leq \alpha_4/\alpha_1$.
This implies that $\alpha_1 (1-p) \leq \alpha_4 p$.  
Hence $\lambda_4 = \alpha_1 p + \alpha_2 p + \alpha_3 p + \alpha_4 p$.
In this case the condition
$\lambda_4'\geq\lambda_4$ is true iff
$r\geq (\alpha_4-\epsilon_3)/(\alpha_1 + \epsilon_1)$.
Also, we have assumed $r\leq\alpha_4/\alpha_1$ and we know
that $\alpha_4\leq\alpha_3\leq\alpha_3+\epsilon_3$ and 
$\alpha_1\geq\alpha_2\geq\alpha_2-\epsilon_1$.  Thus 
$r\leq(\alpha_3+\epsilon_3)/(\alpha_2 - \epsilon_1)$.
Conversely, if both these inequalities hold, then $\lambda_4'\geq\lambda_4$.

For our second case,
suppose $\alpha_4/\alpha_1< r\leq \alpha_3/\alpha_2$.  
(We are only guaranteed that $\alpha_4/\alpha_1\leq\alpha_3/\alpha_2$,
so this case may not occur.)
Then
we get $\lambda_4 = \alpha_1 + \alpha_2 p + \alpha_3 p$ and
$\lambda_4'\geq\lambda_4 \iff
\epsilon_1 + \epsilon_3 p\geq \epsilon_1 p$.
But $p\leq1$ so $\epsilon_1>\epsilon_1 p$ and hence we always have
$\lambda_4'\geq\lambda_4$.
Likewise, since $\alpha_4/\alpha_1< r\leq \alpha_3/\alpha_2$,
both $r\geq (\alpha_4-\epsilon_3)/(\alpha_1 + \epsilon_1)$ and
$r\leq (\alpha_3+\epsilon_3)/(\alpha_2 - \epsilon_1)$ will always hold.

For the final case, suppose $r > \alpha_3/\alpha_2$.  This implies
$\alpha_2 (1-p) > \alpha_3 p$ and thus $\lambda_4 = \alpha_1 + \alpha_2$.
Therefore,
$\lambda_4'\geq\lambda_4 \iff 
r\leq (\alpha_3+\epsilon_3)/(\alpha_2-\epsilon_1)$.
Additionally, 
$r>\alpha_3/\alpha_2\geq (\alpha_4-\epsilon_3)/(\alpha_1+\epsilon_1)$.
Conversely, if both these inequalities hold, then 
$\lambda_4'\geq\lambda_4$.

We have therefore shown the desired equivalence.

\subsection{The fifth sum: $\lambda_5'\geq\lambda_5$ iff
$r\geq (\alpha_4-\epsilon_3)/(\alpha_2 - \epsilon_2)$ and
$r\leq \epsilon_3/\epsilon_2$.}

Fortunately we are back down to considering just two cases.  For the
first, suppose $r\leq \alpha_4/\alpha_2$.   Then
$\alpha_2 (1-p) \leq \alpha_4 p$ and thus 
$\lambda_5 = \alpha_1 + \alpha_2 p + \alpha_3 p + \alpha_4 p$.
Again we can manipulate the inequality to get,
$\lambda_5'\geq\lambda_5 \iff
r \geq (\alpha_4-\epsilon_3)/(\alpha_2 - \epsilon_2)$.
Notice also, that we have assumed $r\leq\alpha_4/\alpha_2$.  Since
$r\geq (\alpha_4-\epsilon_3)/(\alpha_2 - \epsilon_2)$, 
and our preliminary lemma implies
$r\leq \epsilon_3/\epsilon_2$.  
Conversely, if $r\geq (\alpha_4-\epsilon_3)/(\alpha_2 - \epsilon_2)$, then 
$\lambda_5'\geq\lambda_5$.

For the second case $r>\alpha_4/\alpha_2$.  This implies
$\lambda_5 = \alpha_1 + \alpha_2  + \alpha_3 p $ and
$\lambda_5'\geq\lambda_5 \iff
r \leq \epsilon_3/\epsilon_2$.
Also, we have assumed $r>\alpha_4/\alpha_2$.  Since
$r\leq \epsilon_3/\epsilon_2$, 
the preliminary lemma again implies
$r\geq (\alpha_4-\epsilon_3)/(\alpha_2 - \epsilon_2)$. 
Conversely, if $r\geq \epsilon_3/\epsilon_2$, then 
$\lambda_5'\geq\lambda_5$.

We have therefore shown the desired equivalence.  
\vspace{.125in}
\newline
{\bf Remark: }{\it Notice that we require 
$r\geq (\alpha_4-\epsilon_3)/(\alpha_2 - \epsilon_2)$.  Since both the
numerator and denominator are positive, this requires $r$ to be positive.
But we also require $r\leq \epsilon_3/\epsilon_2$.  If $\epsilon_3 = 0$,
this is not possible.  Therefore this step
shows that $\epsilon_3>0$ is necessary for catalysis to occur.  
Hence the weak
inequalities in $(*)$ and $(**)$ may be replaced by strict inequalities.  
(Strictly speaking, we have only shown this for a 2-state catalyst. 
However it is easy to generalize this argument to an
arbitrary $n$-state catalyst by considering the sums of the $n+1$ and 
$3n-1$ largest components of the $4n$-long vectors.)}

\subsection{The sixth sum: $\lambda_6'\geq\lambda_6$ iff
$r\geq (\alpha_4-\epsilon_3)/(\alpha_3 + \epsilon_3)$.}

Again, we consider two cases.  First, suppose $r\leq \alpha_4/\alpha_3$.   
This implies
$\alpha_3 (1-p) \leq \alpha_4 p$ and 
$\lambda_6 = \alpha_1 + \alpha_2 + \alpha_3 p + \alpha_4 p$.
Therefore, $\lambda_6'\geq\lambda_6 \iff
r \geq (\alpha_4-\epsilon_3)/(\alpha_3 + \epsilon_3)$.

Now suppose $r>\alpha_4/\alpha_3$.  This implies
$\lambda_6 = \alpha_1 + \alpha_2  + \alpha_3 $.
Hence
$\lambda_6'\geq\lambda_6 \iff
\alpha_1 + \alpha_2  + \alpha_3 + \epsilon_3 
\geq \alpha_1 + \alpha_2  + \alpha_3$
which is always true since $\epsilon_3\geq0$.  Moreover, since
we are assuming $r>\alpha_4/\alpha_3$, we always have 
$r \geq (\alpha_4-\epsilon_3)/(\alpha_3 + \epsilon_3)$.

We have therefore shown the desired equivalence.

\subsection{The seventh sum: $\lambda_7'\geq\lambda_7$ always holds.}

We know that $\alpha_4 (1-p)$ is the smallest component in $\vec\beta$ so
$\lambda_7 = \alpha_1 + \alpha_2 + \alpha_3 + \alpha_4 p$.  Thus 
$\lambda_7' = \lambda_7 + \epsilon_3 (1-p)$ and we have 
$\lambda_7'\geq\lambda_7$.  This inequality always holds and
there is no further restriction on $r$.

\subsection{The last sum: $\lambda_8'=\lambda_8$ always holds.}

Since $\lambda_8 = \alpha_1 + \alpha_2 + \alpha_3 + \alpha_4 = \lambda_8'$,
this is automatically true and no further restrictions are placed on $r$.

\subsection{Combining the restrictions: $m\leq r \leq M$.}

If we check back through {\bf 3.1}-{\bf 3.12}, we see that if $\vec\beta$ is 
majorized by $\vec\beta'$, then $r$ must be greater than or equal to each 
of the following terms:
$${\alpha_2'\over\alpha_1'},
{\alpha_4'\over\alpha_3'}, 
{\alpha_2-\epsilon_1\over\alpha_1 + \epsilon_1}, 
{\alpha_3+\epsilon_2\over\alpha_1+\epsilon_1}, 
{\epsilon_2\over\epsilon_1},
{\alpha_4-\epsilon_3\over\alpha_1+\epsilon_1},
{\alpha_4-\epsilon_3\over\alpha_2-\epsilon_2},
{\alpha_4-\epsilon_3\over\alpha_3+\epsilon_3}.$$
However $\alpha_2'/\alpha_1'$,  
$(\alpha_3+ \epsilon_2)/(\alpha_1+\epsilon_1)$, and
$(\alpha_4-\epsilon_3)/(\alpha_1+\epsilon_1)$
are less than 
$(\alpha_2-\epsilon_1)/(\alpha_1+\epsilon_1)$; 
and $\alpha_4'/\alpha_3'$ and 
$(\alpha_4-\epsilon_3)/(\alpha_2-\epsilon_2)$ are less
than $(\alpha_4-\epsilon_3)/(\alpha_3+\epsilon_3)$.
Therefore, it is only necessary to require
$$r\geq {\alpha_2-\epsilon_1\over\alpha_1+\epsilon_1},
{\alpha_4 - \epsilon_3\over\alpha_3+\epsilon_3},
{\epsilon_2\over\epsilon_1}$$
In other words, we require $r\geq m$.

Similarly, from {\bf 3.1}-{\bf 3.12}, $r$ must be less than or equal to the 
following terms:
$${\alpha_3'\over\alpha_2'}, 
{\alpha_3+\epsilon_3\over\alpha_2-\epsilon_1},
{\epsilon_3\over\epsilon_2}$$
However $\alpha_3'/\alpha_2'$ is greater than 
$(\alpha_3+\epsilon_3)/(\alpha_2-\epsilon_1)$.  Therefore, it is
only necessary to require
$$r\leq {\alpha_3+\epsilon_3\over\alpha_2-\epsilon_1},
{\epsilon_3\over\epsilon_2}$$
In other words, we require $r\leq M$.
\vspace{.125in}
\newline
{\it We have now completed the first direction of the proof, 
$\vec\beta$ majorized by $\vec\beta'$ implies $m\leq r\leq M$.}

\subsection{The other direction: $m\leq r\leq M$ implies $\vec\beta$ 
is majorized by $\vec\beta'$.}

Assume that $m\leq r\leq M$.  Note that the if and only if statements of 
{\bf 3.5} through {\bf 3.12} depend upon the results derived in {\bf 3.1}
 through {\bf 3.4}.
We will now derive them for this direction.
Since $\alpha_2'/\alpha_1'\leq m\leq r$, we
have $\alpha_2' p \leq \alpha_1' (1-p)$.  Since 
$\alpha_3'/\alpha_2'\geq M\geq r$, we have $\alpha_3' p\geq \alpha_2' (1-p)$.
Since $\alpha_4'/\alpha_3'\leq m\leq r$, we have 
$\alpha_4' p\leq\alpha_3' (1-p)$.  This fixes the ordering of the components
and thus the calculation of the $\lambda_i'$'s done in {\bf 3.4} holds in
this particular case.

We now know the results of {\bf 3.1}-{\bf 3.4} hold.  Also, we know that all of the
inequalities listed in {\bf 3.13} are true.  Thus we may use the
equivalences shown in {\bf 3.5}-{\bf 3.12} and conclude that $\lambda_i'\geq\lambda_i$
for $i=1,\ldots,8$ with equality holding for $i=8$.  Therefore, $\vec\beta$
is majorized by $\vec\beta'$.
\vspace{.125in}
\newline
{\it This completes the proof of the theorem.}

\section{Conclusions}

Note that the arguments of the min and max functions which determine 
$m$ and $M$ are of a special form.  Three of them,
$$ {\alpha_2-\epsilon_1\over\alpha_1+\epsilon_1},
{\alpha_3+\epsilon_3\over\alpha_2-\epsilon_1},
{\alpha_4 - \epsilon_3\over\alpha_3+\epsilon_3}$$
are ratios of the form $\alpha_i'/\alpha_{i+1}'$ with $\epsilon_2$ replaced
by 0.  The other two ratios,
$${\epsilon_2\over\epsilon_1},
{\epsilon_3\over\epsilon_2}$$
compare the sizes of $\epsilon_1$ and $\epsilon_3$ to $\epsilon_2$.

The quantity $\epsilon_2$ can be considered a measure of how much majorization
is violated in the uncatalyzed system. 
Similarly, $\epsilon_1$ and $\epsilon_3$ are the amount of ``slack'' we are 
given to work with in the other components.  If $\epsilon_2$ is large with
respect to $\epsilon_1$, $m$ becomes large; if it is large with respect to
$\epsilon_3$, $M$ becomes small.  We thus require enough ``slack'' on both
ends to make up for the ``bulge'' in the middle.  These notions of ``bulge''
and ``slack'' can be formalized by looking at the areas bounded between the
Lorenz curves generated by $\vec\alpha$ and $\vec\alpha'$.  
See {\bf [4]} for further details.

The other quantities are ratios of the components in the limiting case
where no catalysis is necessary.  This corresponds to looking at the slopes
of the two Lorenz curves.

Finally, we note that this paper deals only with the case of a 4-particle
system evenly divided between two parties and a 2-particle catalyst
similarly divided.  Moreover, we assume that all pieces of this system
have the same Schmidt basis.  This is clearly not the most general case
one could consider.  The next logical generalization would be to analyze
the case of a 4-particle system and $2n$-particle catalyst, all with the
same Schmidt basis.

\section{Some Examples}

We will now return to the examples discussed at the beginning of the 
paper.  Jonathan and Plenio's catalysis example had
$$\begin{array}{rlllr}
|\psi\rangle &= \sqrt{0.4}\,\,|00\rangle &+\sqrt{0.4}\,\,|11\rangle &+
\sqrt{0.1}\,\,|22\rangle &+ \sqrt{0.1}\,\,|33\rangle \\
|\phi\rangle &= \sqrt{0.5}\,\,|00\rangle &+\sqrt{0.25}\,\,|11\rangle &+
\sqrt{0.25}\,\,|22\rangle &+ 0\,\, |33\rangle\\
&&&\\
|\kappa\rangle &= \sqrt{0.6}\,\,|00\rangle &+ \sqrt{0.4}\,\,|11\rangle &&$$
\end{array}$$
This becomes
$\alpha_1 = .4$, $\alpha_2 = .4$, $\alpha_3 = .1$,
$\alpha_4 = .1$, $\epsilon_1 = .1$, $\epsilon_2 = .05$, and
$\epsilon_3 = .1$.  Thus
$$m={\rm max}\left({.3\over.5},
{0\over.2},
{.05\over.1}\right) = {3\over5}$$
and
$$M={\rm min}\left({.2\over.3},
{.1\over.05}\right) = {2\over3}$$
Since $m\leq M$, catalysis is possible and any $|\kappa\rangle$ with 
$3/5\leq r\leq 2/3$ will be a valid catalyst -- in other words,
$3/5\leq p\leq 5/8$.  In this example,  $p=3/5$.

\vspace{.25in}

\noindent In the second example above, we had
$$\begin{array}{rlllr}
|\psi\rangle &= \sqrt{0.45}\,\,|00\rangle &+\sqrt{0.45}\,\,|11\rangle &+
\sqrt{0.05}\,\,|22\rangle &+ \sqrt{0.05}\,\,|33\rangle \\
|\phi\rangle &= \sqrt{0.5}\,\,|00\rangle &+\sqrt{0.35}\,\,|11\rangle &+
\sqrt{0.15}\,\,|22\rangle &+ 0\,\, |33\rangle\\
&&&\\
|\kappa\rangle &= \sqrt{p}\,\,|00\rangle &+ \sqrt{1-p}\,\,|11\rangle &&$$
\end{array}$$ 
This becomes
$\alpha_1 = .45$, $\alpha_2 = .45$, $\alpha_3 = .05$,
$\alpha_4 = .05$, $\epsilon_1 = .05$, $\epsilon_2 = .05$, and
$\epsilon_3 = .05$.  Thus
$$m={\rm max}\left({.4\over.5},
{0\over.1},
{.05\over.05}\right) = 1$$
and
$$M={\rm min}\left({.1\over.4},
{.05\over.05}\right) = {1\over4}$$
Since $m>M$, catalysis is not possible for any value of $p$, 
as we have already seen.

\section{Existence of Specific Values}

We have seen above that the value of $m$ must be positive and the value
of $M$ must be less than 1.  Also, since $M$ is the minimum of two positive
quantities, $M$ must be positive.  Therefore, we can pose the question:

\vspace{.125in}
\noindent {\it Given numbers $m_0$ and $M_0$ with $0<m_0$ and $0<M_0<1$, do 
there exists states $|\psi\rangle$ and $|\phi\rangle$ satisfying $(**)$
for which $m=m_0$ and $M=M_0$?}
\vspace{.125in}

The answer to this question is yes, and we will proceed to give a
construction.  We first consider the case where $m_0\leq1$.  Choose
a positive number $\mu$ with
$$\mu < {\rm min}\left({1\over2}\,{1-M_0\over1+M_0},{1\over2}\,{1-m_0/2\over1+2M_0}\right)$$
and let 
$$a=\left({2\over m_0+2}\right)^2.$$

Let $|\psi\rangle$ and $|\phi\rangle$ be the states given by
$$\begin{array}{rl}
\alpha_1 &= a(1-\mu) \\
\alpha_2 &=a(m_0/2+(m_0+1)\mu) \\
\alpha_3 &= a(m_0/2 - (M_0+1)m_0\mu \\
\alpha_4 &= a(m_0^2/4 + M_0m_0\mu) \\
\alpha_1' &= a \\
\alpha_2' &=am_0/2 \\
\alpha_3' &= am_0/2 \\
\alpha_4' &= am_0^2/4 \\
\end{array}$$

Since $m_0\leq1$, it is clear that the $\alpha'_i$'s are in decreasing
order and one can easily verify that they sum to 1.  The fact that 
$\mu<1/2 (1-m/2)/(1+2M)$ implies that the $\alpha_i$'s are in decreasing
order and it is easy to verify that they also sum to 1.  Computing the
$\epsilon_i$'s, we get
$$\epsilon_1 = \mu a, \quad\epsilon_2 = m_0 \mu a, 
\quad\epsilon_3 = M_0 m_0 \mu a.$$

Performing the calculation of $m$ and $M$, we obtain
$$\begin{array}{rcl}
m &= {\rm max}\left({m_0\over2}(1+2\mu), {m_0\over2}{1\over1-2\mu}, m_0\right)
 &= m_0 \\
M &= {\rm min}\left({1-2\mu\over1+2\mu}, M_0\right) &= M_0. \\
\end{array}$$
Here, the fact that $m_0$ is the largest of the three values is 
straight-forward, while the fact that $M_0$ is the smaller of the
two values follows from the fact that $\mu<1/2 (1-M_0)/(1+M_0)$.
We have therefore produced two states which yield the desired 
$m$ and $M$.

For the case of $m>1$, we set 
$$\mu < {\rm min}\left({1\over2}\,{1-M_0\over1+M_0},{1\over2}\,{(1/2)\over1+2M_0}\right)$$
and define the states $|\phi\rangle$ and $|\psi\rangle$ by 
$$\begin{array}{rl}
\alpha_1 &= a(1-\mu) \\
\alpha_2 &=a(1/2+(m_0+1)\mu) \\
\alpha_3 &= a(1/2 - (M_0+1)m_0\mu \\
\alpha_4 &= a(1/4 + M_0m_0\mu) \\
\alpha_1' &= a \\ 
\alpha_2' &=a/2 \\
\alpha_3' &= a/2 \\ 
\alpha_4' &= a/4 \\
\end{array}$$

Note that if we compute the values of the $\epsilon$'s, they remain unchanged.
Again, the various terms are ordered properly because of the choice of
$\mu$.  Also,
$$\begin{array}{rcl}
m &= {\rm max}\left({1\over2}(1+2\mu), {1\over2}{1\over1-2\mu}, m_0\right)
 &= m_0 \\
M &= {\rm min}\left({1-2\mu\over1+2\mu}, M_0\right) &= M_0. \\
\end{array}$$
Here, the minimality of $M_0$ is the same as before and the maximality
of $m_0$ follows from the fact that $m_0>1$ and the definition of $\mu$.

\vspace{.125in}

We will now provided a concrete example.  Let us choose $m_0=2/3$ and
$M_0=1/3$.  (So we have a case in which catalysis cannot occur.) 
We require
$$0< \mu < {\rm min}\left({1\over2}\,{2/3\over4/3},{1\over2}\,{2/3\over5/3}\right)
 = 1/5$$
so let us arbitrarily choose $\mu = 1/10$.  We set 
$$a=\left({2\over 2/3+2}\right)^2 = \left({3\over4}\right)^2 = {9\over16}.$$
Since $m_0<1$, we set
$$\begin{array}{rrrrl}
\alpha_1 &= 81/160, \quad \alpha_2 &= 45/160,
\quad\alpha_3 &= 22/160, \quad\alpha_4 &= 12/160 \\
\alpha_1' &= 90/160, \quad\alpha_2' &= 30/160,
\quad\alpha_3' &= 30/160, \quad\alpha_4' &= 10/160 \\
\end{array}$$
This yields
$$\epsilon_1 = 9/160, \quad\epsilon_2 = 6/160, 
\quad\epsilon_3 = 2/160.$$
and we have
$$\begin{array}{rll}
m &= {\rm max}\left({36\over90}, {10\over24}, {6\over9}\right)
 &= {2\over3} \\
M &= {\rm min}\left({24\over36}, {2\over6}\right) &= {1\over3} \\
\end{array}$$
as desired.

 \vspace{.25in}
 \centerline{{\bf ACKNOWLEDGMENTS}}
 \vspace{.25in}

Thanks go to Mark Heiligman, Nathan Panike, and Arthur Pittenger for their helpful comments on
the draft of this paper.

%
\end{document}